\documentclass[apjl]{emulateapj}
\usepackage{amsfonts,amsmath,graphicx,natbib,apjfonts,subfigure}

\def\swift{{\it Swift}}
\def\chandra{{\it Chandra}}
\def\cfa{1}
\def\illinois{2}
\def\illinoisbis{3}
\shorttitle{Lack of X-rays from SN\,2014J}
\shortauthors{Margutti et al.}

\begin{document}
\title{No X-rays from the very nearby Type Ia SN\,2014J: constraints on its environment}
%SN che non lascia traccia in X. SN senza ombra. SN che non lascia impronte in X. 

\author{R. Margutti\altaffilmark{\cfa}, J. Parrent\altaffilmark{\cfa},  A. Kamble\altaffilmark{\cfa},
A.~M. Soderberg\altaffilmark{\cfa}, R.~J. Foley\altaffilmark{\illinois,\illinoisbis}, D. Milisavljevic\altaffilmark{\cfa}, M. R. Drout\altaffilmark{\cfa},  R. Kirshner\altaffilmark{\cfa}}

\altaffiltext{\cfa}{Harvard-Smithsonian Center for Astrophysics, 60 Garden St., Cambridge, MA 02138, USA.}
\altaffiltext{\illinois}{Astronomy Department, University of Illinois at Urbana--Champaign, 1002 W.\ Green Street, Urbana, IL 61801 USA.}
\altaffiltext{\illinoisbis}{Department of Physics, University of Illinois Urbana--Champaign, 1110 W.\ Green Street, Urbana, IL 61801 USA.}

\begin{abstract}
Deep X-ray observations of the post-explosion environment around the very 
nearby Type Ia SN\,2014J ($d_{\rm{L}}=3.5\,\rm{Mpc}$) reveal no X-ray emission down 
to a luminosity $L_x<7\times10^{36}\,\rm{erg\,s^{-1}}$ (0.3-10 keV) at $\delta t\sim20\,\rm{days}$ after the explosion.
We interpret this limit in the context of Inverse Compton emission from upscattered 
optical photons by the supernova shock and constrain the pre-explosion mass-loss
rate of the stellar progenitor system to be $\dot M<10^{-9}\,\rm{M_{\sun}y^{-1}}$  (for wind velocity 
$v_w=100\,\rm{km\,s^{-1}}$). Alternatively, the SN shock might be expanding into a uniform
medium with density $n_{\rm{CSM}}<3\,\rm{cm^{-3}}$. These results rule out single-degenerate (SD) 
systems with steady mass-loss  until the 
terminal explosion and constrain the fraction of transferred material lost at the outer Lagrangian point 
to be $\leq1$\%. The allowed progenitors are (i) WD-WD progenitors,
(ii) SD systems with unstable hydrogen burning experiencing recurrent nova eruptions with
recurrence time $t<300$ yrs and (iii) stars where the mass loss ceases before the explosion.

%\textbf{What this limit implies in terms of progenitor systems.}
%Our observations favor double degenerate progenitors or single degenerate systems with $> 500$ yr
%delay between steady (i.e. wind) mass loss and the terminal explosion (for a typical RGS -red giant star-
%wind velocity of 10 km s).
\end{abstract}

%%%%%%%%%%%%%%%%%%%%%%%%%%%%%%%%%%%%%%%%%%%
\section{Introduction}
\label{Sec:Intro}

Type Ia supernovae (SNe) are believed to originate from white dwarfs (WDs) in binary systems. However,
no stellar progenitor has ever been directly identified in pre-explosion images, not even
for the two closest Type Ia SNe discovered in the last 25 years, SN\,2011fe \citep{Li11} and, more recently,
SN\,2014J (\citealt{Kelly14}, \citealt{Goobar14}). As a result,  
the nature of their progenitor system is still a matter of debate (e.g.  \citealt{Howell11}, \citealt{DiStefanoProc}, 
\citealt{Maoz13}). 
Here we present a detailed study of  SN\,2014J at X-ray energies, with the
primary goal to constrain the circumstellar environment of the exploding star.

SN\,2014J was discovered by \cite{Fossey14} in the nearby starburst galaxy M82.
At the distance of $d_{\rm{L}}=3.5\,\rm{Mpc}$ \citep{Dalcanton09,Karachentsev06},
SN\,2014J  is the nearest Type Ia SN discovered in the last three decades
and offers an unprecedented opportunity to study the progenitor
system of thermonuclear stellar explosions. 

Type Ia SNe are believed to originate from the runaway thermonuclear explosion of
a degenerate C/O  stellar core, likely a WD in a binary system 
(\citealt{Hoyle60}, \citealt{Colgate69},
see \citealt{Calder13}, \citealt{Maoz13} and \citealt{Parrent14}
for recent reviews). %The first observational indications of a primary C/O WD were provided by early-time 
%optical observations of SN\,2011fe (\citealt{Nugent11}, \citealt{Bloom12}).
While there is consensus that a white dwarf explodes, the astronomical events that precede the explosion are less clear.
%For 11fe pre-explosion and post-explosion X-ray, radio 
%and optical observations favored a 

Two progenitor channels are mostly favored, involving
non-degenerate and degenerate binary companions, respectively.
%(see however \citealt{Kushnir13}, \citealt{Soker14}, \citealt{Ouyed14}). 
In the first scenario (single-degenerate, SD, hereafter), the WD accretes 
material from the companion, potentially a
giant, sub-giant or main-sequence star  (\citealt{Whelan73}, \citealt{Hillebrandt2000}, \citealt{Nomoto82b}).
Wind from the secondary star provides material to be accreted (symbiotic channel, 
see e.g. \citealt{Patat11}).
Mass can be transferred to the WD via Roche-lobe overflow (RLOF) 
from a hydrogen-rich mass-donor star or He star (\citealt{Nomoto82}, \citealt{vandenHeuvel92},
\citealt{Iben94}, \citealt{Yoon03}).   
The second standard scenario involves two WDs (double-degenerate scenario,
DD, hereafter, \citealt{Iben84}, \citealt{Webbink84}). The explosion
occurs as the two WDs merge as a consequence of the loss of angular 
momentum.% via gravitational waves.

The two classes of progenitor systems are predicted to leave different imprints on the
circumbinary environment. 
For SD progenitor channels the local environment is expected to be either directly enriched by wind from the donor 
star (symbiotic systems) or by non-conservative mass transfer (i.e. by matter that the WD
is unable to accrete). For the DD scenario the general expectation is that of 
a ``cleaner'' environment with density typical of the inter stellar medium (ISM).
Particularly intriguing in this respect  is the evidence for interaction of 
the SN ejecta with circumbinary material as revealed by optical observations of some Type Ia SNe
(see e.g. \citealt{Silverman13}). 

The nearby SN environment can also be revealed by using sensitive X-ray observations
obtained shortly after the explosion (e.g. \citealt{Russell12}, \citealt{Horesh12},
\citealt{Margutti12}). For both the SD and DD scenarios, the X-ray emission 
is powered by the interaction of the
SN shock with the circumburst medium and can thus be used as a probe 
of the local environment (e.g. \citealt{Chevalier06}).
Here we present the results from deep \emph{Chandra} X-ray observations 
of SN\,2014J that enable us to put  stringent constraints on the progenitor
system mass-loss prior to explosion.
X-ray observations are described in Sec. \ref{Sec:Obs}. 
We re-construct the bolometric light-curve of SN\,2014J from
broad-band optical photometry spanning the UV to NIR range in Sec. 
\ref{Sec:bolometric}.  We combine the optical and X-ray observations to 
derive a deep limit to the mass-loss rate of the progenitor system of 
SN\,2014J  in Sec. \ref{Sec:massloss}. 
The mass-loss limit is used to infer the nature of the donor star
in the progenitor system of SN\,2014J in Sec. \ref{Sec:discussion}.
Conclusions are drawn in Sec. \ref{Sec:conclusions}.

%Throughout the paper we use the convention $F_{\nu}(\nu,t)\propto
%\nu^{-\beta}\,t^{-\alpha}$, where the spectral energy index is related
%to the spectral photon index by $\Gamma=1+\beta$.  
Uncertainties are quoted at $1\sigma$ confidence level, unless otherwise noted.
%We employ standard cosmology with $H_{0}=71$ km s$^{-1}$ Mpc$^{-1}$,
%$\Omega_{\Lambda}=0.73$, and $\Omega_{\rm M}=0.27$.
Throughout the paper we use 2014 January 14.72 UT as explosion date of 
SN\,2014J \citep{Zheng14}. The possible presence of a ``dark phase''  (e.g.
\citealt{Piro12,Piro13}) with duration
between hours and a few days between the explosion and the time of the first emitted light has no 
impact on our major conclusions.
%The inferred distance to SN\,2014J is $d_{\rm{L}}=3.5\,\rm{Mpc}$
%\citep{Dalcanton09,Karachentsev06}.

%%%%%%%%%%%%%%%%%%%%%%%%%%%%%%%%%%%%%%%%%%%
\section{Observations and data analysis}
\label{Sec:Obs}

%------------------------------------------------------------------------------------------------------------------
\subsection{\chandra}
\label{SubSec:ChandraObs} 

\begin{figure*}
\vskip -0.0 true cm
\centering
\includegraphics[scale=0.56]{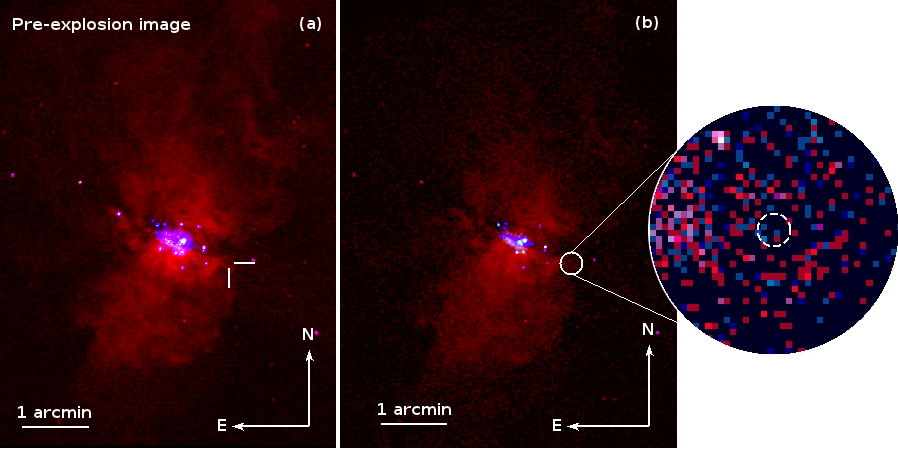}
\caption{False-color pre-explosion (left panel) and post-explosion (central panel)
images of the environment around the Type Ia SN\,2014J obtained with the \emph{Chandra}
X-ray Observatory. Right panel:  zoom-in to the
SN location. The dashed circle has a radius of $0.9''$ around the position of
SN\,2014J (\emph{Chandra} 
PSF at 1.5 keV, 90\% containment). 
No X-ray emission is detected at the position of SN\,2014J, enabling deep limits 
on the mass-loss rate of the stellar progenitor system. The pre-explosion image
combines 237 ks of archival \emph{Chandra} data (PI Strickland). Our post-explosion
observations obtained at $\delta t=20.4\,\rm{days}$ are shown in panel (b) (exposure time of 47 ks, PI Margutti).
Red, green and blue colors refer to soft (0.3-1.4 keV), medium (1.4-3 keV) and hard (3-10 keV) photons, respectively.}
\label{Fig:Chandra}
\end{figure*}

We initiated deep X-ray follow up of SN\,2014J with the \emph{Chandra} X-ray  Observatory
on 2014 February 3, 20:10:39 UT ($\delta t\sim 20.4$ days) under an approved DDT proposal (PI Margutti).
Data have been reduced with the CIAO software package (version 4.6) and corresponding
calibration files. Standard ACIS data filtering has been applied. The total exposure time 
of our observations is $47\,\rm{ks}$. The observations mid-time corresponds to 
$\delta t=20.4\,\rm{days}$ since the explosion. No X-ray source is detected at the
supernova position (Fig. \ref{Fig:Chandra}) with a $3\sigma$ upper limit of $2.1\times10^{-4}\rm{c\,s^{-1}}$
in the 0.3-10 keV energy band, which implies an absorbed flux limit of 
$F_x<2.6\times10^{-15}\,\rm{(erg\,s^{-1}cm^{-2})}$ for an assumed $F_{\nu}\propto \nu^{-1}$
spectrum. The count-rate limit is calculated as a $3\sigma$ fluctuation from the
local background assuming Poisson statistics as appropriate in the 
regime of low-count statistics. The presence of diffuse soft X-ray emission 
from the host galaxy (Fig. \ref{Fig:Chandra}, left panel) prevents us from reaching 
deeper limits. %Restricting our analysis to the  nn nn keV energy band

The Galactic neutral hydrogen column density in the direction of SN\,2014J is 
$\rm{NH}_{\rm{MW}}=5.1\times10^{20}\,\rm{cm^{-2}}$ \citep{Kalberla05}.
The estimate of the intrinsic hydrogen column ($\rm{NH}_{\rm{int}}$) is more uncertain. 
Observations of SN\,2014J at optical wavelengths point to a large local extinction (see
Sec. \ref{Sec:bolometric}) corresponding to $A_V=1.7\pm 0.2\,\rm{mag}$
\citep{Goobar14}. A more recent study by \cite{Amanullah14} find consistent results around 
maximum light: $A_V=1.9\pm 0.1\,\rm{mag}$.
Assuming a Galactic dust-to-gas ratio, this would 
imply $\rm{NH}_{\rm{int}}\sim 4\times 10^{21}\,\rm{cm^{-2}}$ (\citealt{Predehl95,Watson11}). However, the low
value inferred for the total-to-selective extinction $R_V<2$ \citep{Goobar14, Amanullah14}
suggests the presence of smaller dust grains, more similar to the
grain size distribution of the Small Magellanic Cloud (SMC). 
For an SMC-like dust-to-gas ratio \citep{Martin89}, the inferred hydrogen column is
$\rm{NH}_{\rm{int}}\sim 7\times 10^{21}\,\rm{cm^{-2}}$. Gamma-Ray Burst (GRB)
host galaxies sample instead the high end of the gas-to-dust ratio distribution
\citep{Schady10}. For a GRB-like environment the observed optical extinction
would imply $\rm{NH}_{\rm{int}}\sim 10^{22}\,\rm{cm^{-2}}$.
Based on these findings, in the following we calculate our  flux limits 
and results for a fiducial range of intrinsic hydrogen 
column density values: 
$4\times 10^{21}\,\rm{cm^{-2}}<\rm{NH}_{\rm{int}}<10^{22}\,\rm{cm^{-2}}$.
We adopt $\rm{NH}_{\rm{int}}\sim 7\times 10^{21}\,\rm{cm^{-2}}$
as ``central value'' in our calculations.\footnote{Based on pre-explosion \emph{Chandra}
observations, \cite{Nielsen14} infer $\rm{NH}_{\rm{int}}=(8.6\pm0.4)
\times 10^{21}\,\rm{cm^{-2}}$.  However, the actual X-ray absorption affecting SN\,2014J
can be either smaller or larger, depending if SN\,2014J exploded in front of some of the
material responsible for the intrinsic hydrogen column in pre-explosion
images or if instead SN\,2014J illuminated additional material which was not
contributing to the measured $\rm{NH}_{\rm{int}}$. }

Based on these values, the unabsorbed flux limit is 
$F_x^{3\sigma}<5.0\times10^{-15}\,\rm{erg\,s^{-1}cm^{-2}}$
($\rm{NH}_{\rm{int}}=7\times 10^{21}\,\rm{cm^{-2}}$,  0.3-10 keV energy band), 
with $(4.0<F_x^{3\sigma}<6.0)\times10^{-15}\,\rm{erg\,s^{-1}cm^{-2}}$
for the range of $\rm{NH}_{\rm{int}}$ values above. At the distance of 3.5 Mpc,
the corresponding luminosity limit is $L_x^{3\sigma}<7.2\times10^{36}\,\rm{erg\,s^{-1}}$,
with $(5.7<L_x^{3\sigma}<8.7)\times10^{36}\,\rm{erg\,s^{-1}}$.

\subsection{\swift-XRT}
\label{SubSec:XRTObs}

The \emph{Swift} \citep{Gehrels04} X-Ray Telescope (XRT, \citealt{Burrows05})
started observing SN\,2014J on 2014 January 22, 10:13:52 UT ($\delta t\sim 8$ days, 
PIs Ofek, Brown, Markwardt, Barthelmy).
XRT data have been analyzed using  HEASOFT (v6.15) and corresponding calibration files.
Standard filtering and screening criteria have been applied.  We find 
no evidence for a point-like X-ray source at the position of SN\,2014J.
Using observations acquired at $\delta t<20$ days (total exposure time of 114 ks),
the $3\,\sigma$ count-rate limit is $2.4\times10^{-3}\,\rm{c\,s^{-1}}$ in the
0.3-10 keV energy range.  The unabsorbed flux limit is 
$F_x^{3\sigma}<2.4\times 10^{-13}\,\rm{erg\,s^{-1}cm^{-2}}$  
($\rm{NH}_{\rm{int}}=7\times 10^{21}\,\rm{cm^{-2}}$), corresponding
to $L_x^{3\sigma}<3.5\times 10^{38}\,\rm{erg\,s^{-1}}$. Restricting our analysis
to $7\lesssim \delta t\lesssim 15$ days, when X-ray Inverse Compton emission is expected
to peak (Fig. \ref{Fig:IC}), we find a count-rate limit of $5.4\times10^{-3}\,\rm{c\,s^{-1}}$
(exposure time of 27 ks, 0.3-10 keV), corresponding to $F_x^{3\sigma}<5.5\times 10^{-13}\,\rm{erg\,s^{-1}cm^{-2}}$ 
with $L_x^{3\sigma}<8.0\times 10^{38}\,\rm{erg\,s^{-1}}$.

%%%%%%%%%%%%%%%%%%%%%%%%%%%%%%%%%%%%%%%%%%
\section{The Bolometric Luminosity of SN\,2014J}
\label{Sec:bolometric}

\begin{figure}
\vskip -0.0 true cm
\centering
\includegraphics[scale=0.75]{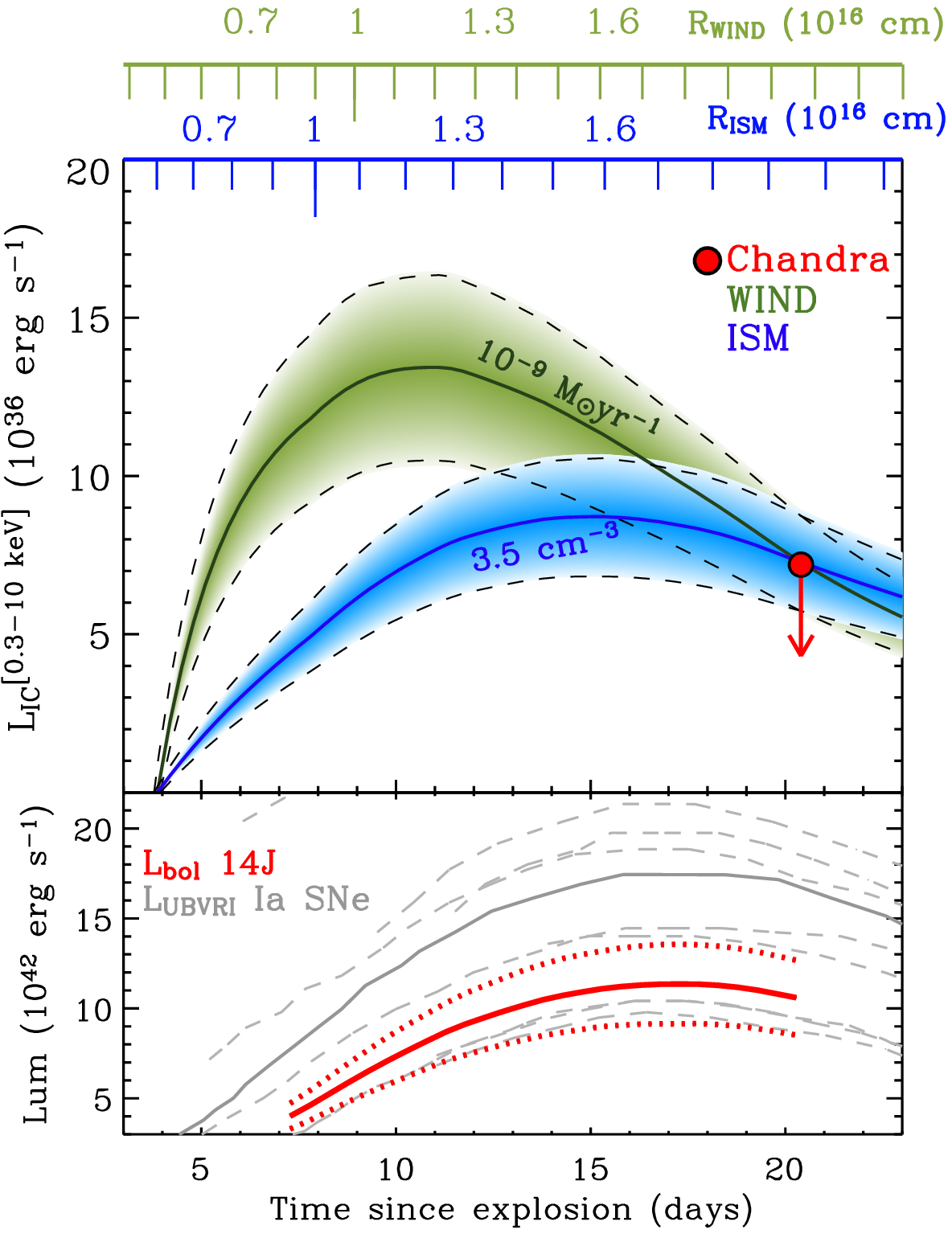}
\caption{\emph{Upper panel:} Inverse Compton X-ray luminosity expected 
in the case of wind (green solid line) and ISM (blue solid line) environments.
The deep X-ray limit obtained with \emph{Chandra} constrains 
$\dot M<1.2\times10^{-9}\,\rm{M_{\sun}yr^{-1}}$  (wind) or $n_{\rm{CSM}}<3.5\,\rm{cm^{-3}}$
(ISM). The shaded areas mark the range of allowed limits, derived by 
conservatively accounting for 
the uncertainty on $L_{\rm{bol}}$ and $\rm{NH}_{\rm{int}}$:
$\dot M=(0.7-2.5)\times10^{-9}\,\rm{M_{\sun}yr^{-1}}$  (wind) and 
$n_{\rm{CSM}}=(1.5-8.0)\,\rm{cm^{-3}}$ (ISM).
\emph{Lower panel}: optical bolometric luminosity of SN\,2014J (thick line) and associated
uncertainty (dotted lines). A sample of UBVRI light-curves of Type Ia SNe including
SNe 1989B, 1991T, 1992A, 1992bc, 1992bo,1994D, 1994ae,1995D, 2011fe (thick solid line), 2012cg, 
is also shown for comparison (from \citealt{Munari13}).}
\label{Fig:IC}
\end{figure}

The line of sight toward SN\,2014J is heavily reddened by dust in the host galaxy. The Galactic
reddening in the direction of the SN is $E(B-V)_{\rm{MW}}=0.06\,\rm{mag}$ \citep{Dalcanton09}.
The analysis of the spectra and broad-band photometry of SN\,2014J by \cite{Goobar14}
indicates a local color excess $E(B-V)_{\rm{host}}\sim1.2\,\rm{mag}$
and a low value of total-to-selective extinction $R_V<2$. These results are consistent with 
earlier reports by \cite{Cox14,Patat14}.
% and have been confirmed by a later
%study by \cite{Amanullah14}.
In the remaining of the paper we follow \cite{Goobar14} 
and use $E(B-V)_{\rm{host}}=1.22\pm0.05\,\rm{mag}$ with $R_V^{\rm{host}}=1.40\pm 0.15\,\rm{mag}$.\footnote{The
properties of diffuse interstellar bands (DIBs) in the spectra of SN\,2014J 
would indicate a higher  local extinction corresponding to $A_V^{\rm{host}}=2.5\pm0.1\,\rm{mag}$ \citep{Goobar14}.
A later study by \cite{Amanullah14} finds $A_V^{\rm{host}}=1.9\pm0.1\,\rm{mag}$.
The parameters adopted above imply instead $A_V^{\rm{host}}=1.7\pm0.2\,\rm{mag}$ 
which leads to more conservative limits to the mass-loss rate of the progenitor system
derived in Sec. \ref{Sec:massloss}. } 
The uncertainty on the local extinction parameters is consistently propagated into our final
bolometric luminosity (Fig. \ref{Fig:IC}, lower panel). 

To derive the bolometric luminosity of SN\,2014J we start from the
UBVRYJHK and $iz$ photometry published by \cite{Goobar14}. We complement this photometric
data set with additional JHK photometry from \cite{Venkataraman14,Srivastav14,Richardson14}
and obtain extinction corrected flux densities applying a \cite{Cardelli89} extinction law
with the color excess and the total-to-selective extinction values as above (in addition
to the Galactic correction). A first, pseudo-bolometric light-curve of SN\,2014J is then
obtained by integrating the extinction corrected flux densities from the U band 
($\lambda\sim3650$ \AA) to the K band ($\lambda\sim2.2\,\rm{\mu m}$).
From well monitored ``normal'' Type Ia SNe like SN\,2011fe
we estimate that the amount of flux redward the K band is $\lesssim 5\%$ at $\delta t\lesssim 20\,\rm{days}$
since the explosion. 

An UV photometric campaign (PI Brown) has also been carried out with the \emph{Swift}
\citep{Gehrels04} UV Optical Telescope (UVOT, \citealt{Roming05}). However, the large 
local extinction makes any extinction correction to the UV photometry 
extremely uncertain. For this reason we employ a different approach.
%The UV properties of Type Ia SNe display a high degree of homogeneity within the
%``normal'' subclass \citep{Milne10}. 
We estimate the
time-varying amount of flux emitted blueward the U band by SN\,2014J using 
the well-monitored, minimally reddened, normal Type Ia SN\,2011fe as template 
\citep{Brown12,Margutti12}. At the time of the \emph{Chandra}
observation ($\delta t=20.4\,\rm{days}$) the UV flux  represents
$(13\pm5)\%$ of the bolometric flux.

Figure \ref{Fig:IC}, lower panel,  shows the final bolometric light-curve of SN\,2014J.
The displayed uncertainty is driven by the inaccurate knowledge of the local extinction
correction. In this work we use the estimates of $A_V^{\rm{host}}$ and $R_V^{\rm{host}}$ 
that appeared in the literature so far. It is possible that later estimates will somewhat deviate 
from the numbers assumed here. We emphasize that a wrong choice of extinction parameters
would have no impact on our conclusions as long as it does not lead to 
an overestimate of the bolometric luminosity of SN\,2014J
at $\sim20$ days. The comparison with the UBVRI light-curves of a sample of Type Ia SNe 
in Fig. \ref{Fig:IC} suggests that we might have instead \emph{underestimated} the bolometric
luminosity of  SN\,2014J. The study by \cite{Amanullah14} finds that SN\,2014J is indeed similar
to SN\,2011fe once corrected for the extinction. If this is true, then our limits on the density of the explosion local 
environment should be considered conservative, thus strengthening our conclusions
on the progenitor system of SN\,2014J of Sec. \ref{Sec:discussion}.  
A larger $L_{\rm{bol}}$ would require a smaller environment density in order to avoid detection 
of X-rays through Inverse Compton scattering.

% dense intervening material 
% is indicative of unusually large extinction by dust in the line of sight.
%%%%%%%%%%%%%%%%%%%%%%%%%%%%%%%%%%%%%%%%%%%
\section{Constraints on the progenitor system mass-loss rate}
\label{Sec:massloss}

Deep X-ray observations constrain the density of the 
explosion circumstellar environment, previously shaped by the mass-loss of the progenitor system.
For hydrogen-stripped progenitors in low density environments, at $\delta t \lesssim 40$ days,
the X-ray emission is dominated by Inverse Compton (IC) scattering of 
photospheric optical photons by relativistic electrons accelerated by the SN
shock \citep{Chevalier06}. Following the generalized formalism developed in
\cite{Margutti12}, the IC luminosity  directly tracks the SN optical 
luminosity ($L_{\rm{IC}}\propto L_{\rm{bol}}$) and further depends on the density structure of the SN ejecta
$\rho_{\rm{SN}}$, the structure of the circumstellar medium $\rho_{\rm{CSM}}$ and the details of the electron
distribution responsible for the up-scattering of the optical photons
to X-ray energies. The dynamical evolution of the shockwave is treated self-consistently.
Finally, since $L_{\rm{IC}}\propto L_{\rm{bol}}$, any uncertainty on the estimate of the 
SN distance from the observer would equally affect $L_{\rm{IC}}$ and $L_{\rm{bol}}$,
and thus has no impact on the limits we derive on the environment density.

In the following we assume the SN outer density structure to scale
as $\rho_{\rm{SN}}\propto R^{-n}$ with $n\sim 10$, as found for SNe arising from compact progenitors 
(e.g. \citealt{Matzner99}). Electrons are assumed to be accelerated in a power-law
distribution $n(\gamma)\propto \gamma^{-p}$ with index $p=3$, as supported
by radio observations of SN shocks in Type Ib/c explosions (e.g. \citealt{Soderberg06}).
The fraction of post-shock energy density in relativistic electrons is
$\epsilon_e=0.1$ \citep{Chevalier06}. The environment density limits calculated below
scale as $(\epsilon_e/0.1)^{-2}$ (\citealt{Margutti12}, their Appendix A).

We use the bolometric luminosity light-curve derived in Sec. \ref{Sec:bolometric}
to constrain the density of the environment around SN\,2014J in the case of
(i) a wind-like CSM ($\rho_{\rm{CSM}}\propto R^{-2}$) and (ii) an ISM-like CSM
($\rho_{\rm{CSM}}=const$). A star which has been losing material at constant rate $\dot M$
gives rise to a wind-like CSM. A ``wind medium'' is the simplest expectation in the case of
SD progenitor models. DD progenitor systems would be instead consistent
with a ``cleaner environment'', and, potentially, with an ISM-like CSM.
%In both cases we combine the uncertainty on the bolometric luminosity 
%(dotted lines in Fig. \ref{Fig:IC}, lower panel) with the X-ray luminosity limits of Sec. \ref{SubSec:ChandraObs}
%and determine the range of environment density values that would be consistent
%with our measurements.  
The final result is shown in Fig. \ref{Fig:IC}, upper panel.

For a wind-like medium, $\rho_{\rm{CSM}}=\dot M/(4\pi R^2 v_w)$, where
$\dot M$ is the progenitor pre-explosion mass-loss rate and $v_w$ is the wind 
velocity. The X-ray non-detection by \emph{Chandra} at $\delta t=20.4\,\rm{days}$ 
constrains the progenitor mass-loss rate $\dot M<1.2\times10^{-9}\,\rm{M_{\sun}yr^{-1}}$  
for $v_w=100\,\rm{km\,s^{-1}}$. This limit is obtained by using our fiducial $L_{\rm{bol}}$ (thick
line in Fig. \ref{Fig:IC}, lower panel) and $L_x^{3\sigma}<7.2\times 10^{36}\,\rm{erg\,s^{-1}}$
(Sec. \ref{SubSec:ChandraObs}).
Conservatively accounting for the uncertainties affecting our estimates of the bolometric
optical luminosity and intrinsic neutral hydrogen absorption column, the range of allowed limits to 
the progenitor mass-loss rate is $(0.7-2.5)\times10^{-9}\,\rm{M_{\sun}yr^{-1}}$ 
(shaded areas in Fig. \ref{Fig:IC}, upper panel). In a wind-like scenario
$\dot M/v_{w}\propto(1/L_{\rm{bol}})^{1/0.64}$.
For an ISM-like medium we  find $n_{\rm{CSM}}<3.5\,\rm{cm^{-3}}$. The range of allowed
limits is $(1.5-8.0)\,\rm{cm^{-3}}$ and the particle density scales as $(1/L_{\rm{bol}})^{0.5}$.
% report below the CSM density limits for our fiducial $L_{\rm{bol}}$ (thick
%line in Fig. \ref{Fig:IC}, lower panel) and $L_x^{3\sigma}<7.2\times 10^{36}\,\rm{erg\,s^{-1}}$
%derived from \emph{Chandra} observations (Sec. \ref{SubSec:ChandraObs}),
%together with the range of limits to $\rho_{\rm{CSM}}$ derived as follows.
%The most conservative limit to the environment density is obtained by combining
%the least luminous $L_{\rm{bol}}$ with our shallowest X-ray flux limit.
%By using the most luminous $L_{\rm{bol}}$ of Sec. \ref{Sec:bolometric} and the deepest 
%X-ray flux limit allowed by our analysis of Sec. \ref{SubSec:ChandraObs},
%we instead place the most stringent constraints to the CSM density.

\emph{Swift}-XRT observations have been acquired starting from $\delta t=8$ days
and cover the time interval when IC X-ray emission peaks. From Figure
\ref{Fig:IC}, $L_{\rm{IC}}^{\rm{peak}}<2\times 10^{37}\,\rm{erg\,s^{-1}}$,
consistent with the X-ray limit derived from XRT observations
$L_x^{3\sigma}<8.0\times 10^{38}\,\rm{erg\,s^{-1}}$ (Sec. \ref{SubSec:XRTObs}).

%%%%%%%%%%%%%%%%%%%%%%%%%%%%%%%%%%%%%%%%%%%
\section{Discussion}
\label{Sec:discussion}

\begin{figure}
\vskip -0.0 true cm
\centering
\includegraphics[scale=1.0]{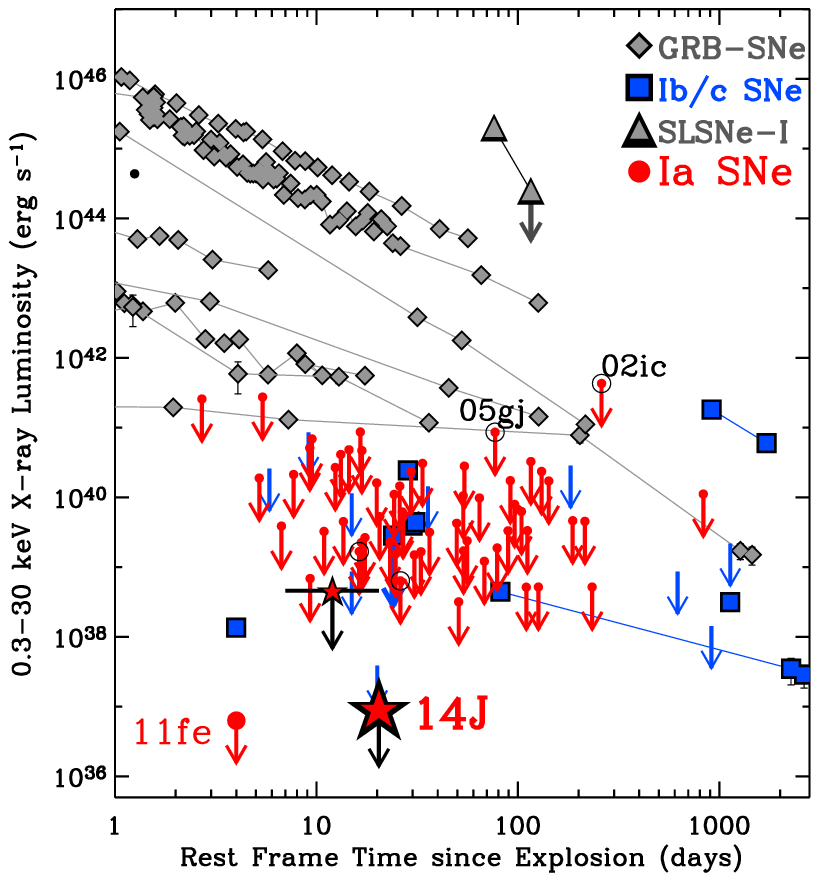}
\caption{Type-I SN explosions in the X-ray phase space, including GRB-SNe
\citep{Margutti13}, ordinary Type Ib/c SNe (\citealt{Margutti14b} and references therein), 
super-luminous hydrogen poor SNe (SLSNe-I, \citealt{Levan13}) and Type-Ia SNe
(\citealt{Russell12}; \citealt{Schlegel93}; \citealt{Hughes07}). The emission from 
SN\,2011fe \citep{Margutti12} and SN\,2014J (stars) is much weaker than any hydrogen-poor core-collapse
SN ever detected in the X-rays, and represent the deepest limit on the
X-ray emission of a Type-I SN to date. Open black symbols mark Type-Ia SNe with 
signs of CSM interaction in our sample (i.e. SNe 2002ic, 2005gj, 2006X, 2009ig).}
\label{Fig:Xrays}
\end{figure}

\begin{figure*}
\vskip -0.0 true cm
\centering
\includegraphics[scale=0.6]{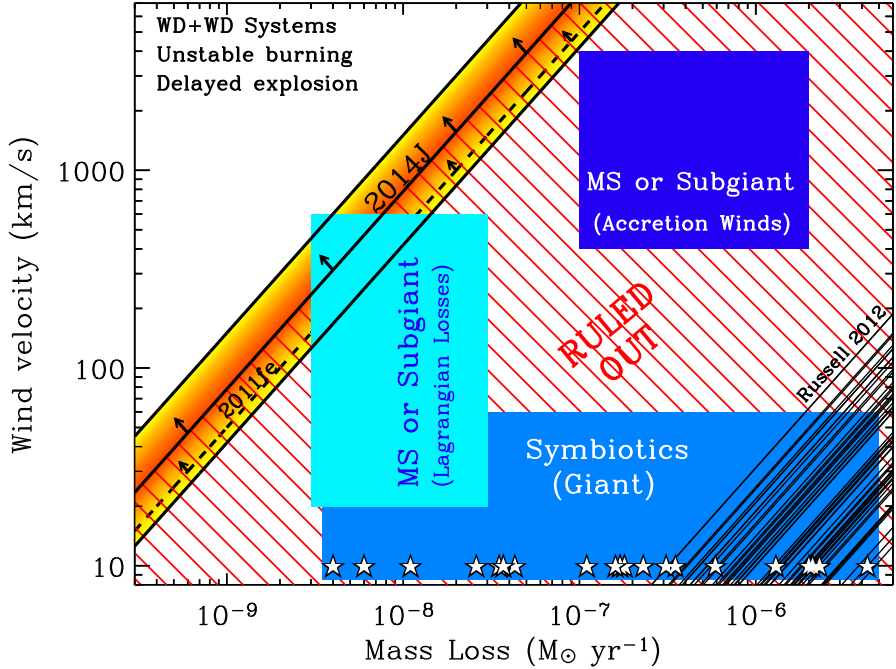}
\caption{Wind velocity vs. mass-loss phase space. The deep limits we obtain for SN\,2014J
(colored area) rule out most of the parameter phase space associated with SD models with steady mass-loss
to the environment. 
The limit obtained from the X-ray observations of SN\,2011fe (dashed line, from \citealt{Margutti12}) and the set of limits
obtained by \cite{Russell12} on a sample of Type Ia SNe (solid black lines) are also shown for comparison. White stars: 
measured mass-loss
of symbiotic systems in our Galaxy for an assumed $v_w=10\,\rm{km\,s^{-1}}$ \citep{Seaquist90}.}
\label{Fig:SDphasespace}
\end{figure*}

Deep X-ray observations obtained around optical 
maximum light of the nearby SN\,2014J allowed us to obtain the most constraining limits on the
environment density around a Type Ia SN explosion. 
With SN\,2011fe \citep{Margutti12} and SN\,2014J (this work), we have probed X-ray luminosities
which are a factor $\sim 100$ deeper than previously obtained limits (Fig. \ref{Fig:Xrays}),
thus sampling a new territory in the X-ray luminosity vs. time phase space. 
However, no X-ray emission is detected in either case, in sharp contrast with 
hydrogen-stripped core-collapse explosions (squares, diamonds and triangles in Fig. \ref{Fig:Xrays}).

These observations directly constraint the density of material in the SN immediate environment, 
shaped by the progenitor system before the terminal explosion. Irrespective of the
assumed circumburst density profile, the deep X-ray limit implies a ``clean'' 
environment at distances $R\lesssim 10^{16}$ cm  from the explosion center of SN\,2014J. 
We interpret this finding in the context of different progenitor configurations, both for SD and DD
models.
%X-ray observations directly constraint the properties of the material with which the
%SN shock is interacting. In particular, the deep null detection can either be the consequence 
%of a very low density environment (significantly more dilute than in the case of core-collapse
%SNe) or might be related to a reduced efficiency of Type Ia SN shocks in accelerating 
%electrons into a non-thermal distribution extending to relativistic speeds, 
%even in the presence of a dense wind. 
%The latter condition naturally arises if the SN shock is interacting with a highly 
%magnetized medium with magnetization $\sigma\propto B^2/n_{\rm{CSM}}>10^{-4}$.
%---------------------------------------------------------------------------
\subsection{SD with quasi-steady mass-loss}

WDs around the Chandrasekhar mass $M_{\rm{Ch}}$ accreting at rate $M_{\rm{acc}}\gtrsim3\times 10^{-7}
\,\rm{M_{\sun}yr^{-1}}$ undergo steady hydrogen burning (e.g. \citealt{Iben82}, \citealt{Nomoto82b}, 
\citealt{Prialnik95}, \citealt{Shen07}). In this regime the WD accretes and retains matter. However, the system
also loses material to the environment in three different ways: 
(i) wind from the donor star; (ii) non-conservative mass
transfer through Roche Lobe Overflow (RLOF); (iii) optically thick winds from the WD.

In symbiotic systems the wind from the giant companion star dominates the mass-loss,
with typical rates $\dot M=5\times10^{-9}-5\times10^{-6}\,\rm{M_{\sun}yr^{-1}}$ and
wind velocities $v_{w}<100\,\rm{km\,s^{-1}}$ (\citealt{Seaquist90}, \citealt{Patat11},
\citealt{Chen11}). Our limit on the environment density around SN\,2014J 
strongly argues against this progenitor scenario (Fig. \ref{Fig:SDphasespace}).
From the analysis of the complete set of pre-explosion images (near UV to NIR) at the SN
site acquired with the Hubble Space Telescope (HST), \cite{Kelly14} exclude 
SD progenitor systems with a bright red giant donor star. Consistent with these results,
the mass-loss limit we derived from deep X-ray observations of the SN environment
firmly and independently rules out this possibility and extends the conclusion
to the entire class of red giant secondary stars.
 
For systems harboring a main sequence, subgiant or helium star, the dominant
source of mass-loss to the surroundings is through non conservative mass
transfer from the donor to the primary, accreting star. In this scenario the secondary star
fills its Roche lobe and some of the transferred 
material is lost to the environment at the outer Lagrangian point: $\dot M_{\rm{trans}}=
\dot M_{\rm{acc}}+\dot M_{\rm{lost}}$. We use the orbital velocity 
$v\sim$ a few $100\,\rm{km\,s^{-1}}$ (or lower)  as typical velocity of the ejected material.
Observations of WDs in this regime indicate velocities up to $\sim600\,\rm{km\,s^{-1}}$
which we use in Fig. \ref{Fig:SDphasespace} (e.g. \citealt{Deufel99}).
The efficiency of the accretion process  ($\dot M_{\rm{acc}}/\dot M_{\rm{trans}}$) is poorly constrained. Our X-ray limit implies
a very high efficiency of $\gtrsim 99\%$ (light-blue
region in Fig. \ref{Fig:SDphasespace}) to avoid detectable signs of interaction
with the lost material. A similar value was found for SN\,2011fe using deep X-ray and radio
observations (\citealt{Margutti12}, \citealt{Chomiuk12}). In this context \cite{Panagia06} 
put a less constraining limit of $>60-80\%$ on the efficiency of the same process 
using radio observations of a larger sample of 27 SNe Ia.

At high mass transfer rates, optically thick winds developing
at the WD surface are expected to self-regulate the mass accretion to a critical 
value $\dot M_{\rm{acc}}\sim 7\times 10^{-7}\,\rm{M_{\sun}\,yr^{-1}}$, the exact value
depending on the hydrogen mass fraction and WD mass
(\citealt{Hachisu99}, \citealt{Han04}, \citealt{Shen07}). Our analysis argues against this
progenitor scenario, even in the case of fast  outflows with 
$v_{\rm{w}}\sim$ a few $1000\,\rm{km\,s^{-1}}$ (dark blue region in Fig. \ref{Fig:SDphasespace}).

Finally, thermonuclear burning of hydrogen rich material on the surface of a WD is
expected to generate super-soft X-ray emission, which should be thus detectable in 
pre-explosion images at the SN site. Consistent with our findings above, \cite{Nielsen14}
found no evidence for a super-soft X-ray source (SSS) at the
position of SN\,2014J using deep \emph{Chandra} X-ray observations spanning the
time range $1\lesssim t \lesssim 14$ yr before explosion. These observations allowed
\cite{Nielsen14} to rule out single-degenerate systems with high effective temperature
of the super-soft emission $kT_{\rm{eff}}>80\,\rm{eV}$.
%---------------------------------------------------------------------------
\subsection{SD with non-steady mass-loss}

\begin{figure}
\vskip -0.0 true cm
\centering
\includegraphics[scale=0.4]{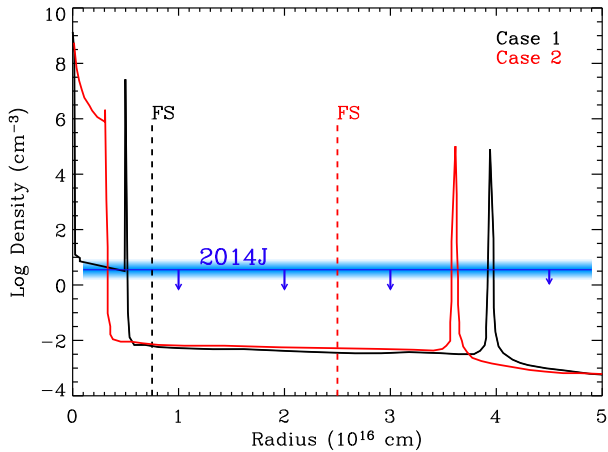}
\caption{CSM density structure around a symbiotic progenitor system (WD plus red giant star) undergoing repeated 
nova explosions with recurrence time $\Delta t=100$ yr  (from \citealt{Dimitriadis14}). 
Case 1: the SN occurs just after the last nova explosion. 
Case 2: the SN explodes $100$ yrs after the last nova shell ejection  and the nova cavity has been 
partially refilled by the wind of the red giant star with $\dot M=10^{-6}\,\rm{M_{\sun}yr^{-1}}$
and $v_w=10\,\rm{km\,s^{-1}}$. 
Dashed vertical lines: radius of the SN forward shock
$\sim20$ days after the explosion (time of our \emph{Chandra} observations). The CSM densities
of the environment sampled by the SN shock at this time are a factor $\sim100$ smaller than our limit.}
\label{Fig:novashell}
\end{figure}

Non-steady mass enrichment of the progenitor surroundings in the years
preceding the SN can create a low-density cavity around the explosion,
which would naturally explain our X-ray null-detection.
The evacuated region around the progenitor  system can either be (i) the consequence 
of the cessation of mass-loss from the companion star or  (ii) the result of 
repeated nova shell ejections by the WD.

\emph{Recurrent novae.} A WD near the Chandrasekhar mass ($M\ge 1.3\,\rm{M_{\sun}}$) 
accreting at $\dot M_{\rm{acc}}\sim
(0.1-3)\times 10^{-7}\,\rm{M_{\sun}yr^{-1}}$ experiences repeated nova
explosions as a result of unsteady hydrogen burning on its surface
(e.g. \citealt{Iben82}, \citealt{Starrfield85}, \citealt{Livio92}, \citealt{Yaron05}).
Shell ejections associated with the nova outbursts evacuate a region 
around the progenitor system by sweeping up the wind from the companion
star. The result is a complex CSM structure,
shaped by the fast nova shells and by the slower wind from the donor star.
This scenario has been invoked  to explain the evidence
for interaction in the spectra of some Type Ia SNe (e.g. SN\,2002ic, \citealt{Wood06}).
The main outcome of this process is the presence of regions with very low
density, reaching $n_{\rm{CSM}}\sim10^{-3}-1\,\rm{cm^{-3}}$, in the
proximity of the explosion center even in the case
of very powerful winds from the donor star (Fig. \ref{Fig:novashell}).

The immediate SN environment critically depends on the time of the SN 
explosion with respect to the time of the last nova shell ejection, as illustrated by
Fig. \ref{Fig:novashell}. Figure  \ref{Fig:novashell} shows the results
from the simulations by \cite{Dimitriadis14} for a symbiotic progenitor system consisting 
of a WD plus red giant star losing material with rate 
$\dot M =10^{-6}\,\rm{M_{\sun}yr^{-1}}$ and wind velocity 
$v_w=10\,\rm{km\,s^{-1}}$. The nova recurrence time scale in this simulations is $\Delta t=100$ yrs.

From Fig. \ref{Fig:novashell} it is clear that our observations obtained 
at $\delta t\sim20$ days do not probe the range of densities associated with nova cavities.\footnote{However,
according to \cite{Dimitriadis14}, their Fig. 12, we should have been able to detect
X-ray emission at the level of $L_x\ge 5\times 10^{37}\,\rm{erg\,s^{-1}}$ at $\delta t\sim
20$ days if the X-rays are of thermal origin. Such emission is clearly ruled out by our  
observations.}
However, \emph{if} the progenitor of SN\,2014J experienced recurrent nova outbursts,
our X-ray null-detection implies that: (i) a nova shell must have been able to clear out 
the environment at $R\sim10^{16}\,\rm{cm}$; (ii) the wind from the donor star has not yet been
able to refill this volume by the time of the SN explosion.
 
The size of the cavity cleared out by a nova shell depends on the shell
dynamics (e.g. \citealt{Moore12}). 
For typical shell ejection velocities $v_{\rm{sh}}=1000-4000\,\rm{km\,s^{-1}}$
and mass $M_{\rm{sh}}=10^{-7}-10^{-5}\,\rm{M_{\sun}}$ (\citealt{Yaron05})
expanding into a medium enriched by the secondary star mass-loss with rate $\dot M=10^{-7}-10^{-6}
\,\rm{M_{\sun}y^{-1}}$, the shell rapidly evolves from free expansion to the Sedov-Taylor
phase on a time-scale of $t_{\rm{ST}}\sim$ a few days. It then transitions to the
snowplow phase at  $t_{\rm{SP}}\sim$ a few months. Adopting $v_{\rm{sh}}=4000\,\rm{km\,s^{-1}}$,
$t_{\rm{ST}}=2$ days, $t_{\rm{SP}}=2$ months as typical parameters as indicated by observations 
of the recurrent nova RS Oph (e.g. \citealt{Bode87}, \citealt{Mason87}, \citealt{Sokoloski06}), 
a nova shell would reach $R\sim10^{16}\,\rm{cm}$ on a time scale of $\sim40$ yrs.
For a wind environment 
$R\propto t^{2/3}$ during the Sedov-Taylor phase and $R\propto t^{1/2}$ during the snowplow phase.
The slower wind from the donor star would need $\Delta t_{\rm{refill}}\sim300$ yrs 
to refill this region with new material (for $v_{w}=10\,\rm{km\,s^{-1}}$).

These results indicate that \emph{if} the recurrent nova scenario applies to
SN\,2014J, then (i) at least one nova shell has been ejected at $t\ge 40$ yrs before the
SN; (ii) the recurrence time scale of nova outbursts is shorter than $300$ yrs (consistent
with the time scales observed in recurrent novae, the shortest known time between outbursts being
$\sim 1$ yr at the time of writing, \citealt{Tang14}). 
%These numbers should be considered 
To this respect it is worth mentioning that a complementary search for 
nova outbursts in archival optical images revealed no evidence for
nova explosions at the location of SN\,2014J in the $\sim1500$ days
before the SN \citep{Goobar14}. No evidence for interaction has been found   
for SN\,2014J at $\delta t\leq20$ days since the explosion, 
either in the form of light-curve re-brightenings, $\rm{H\alpha}$ emission
or time-variable Na D features  (\citealt{Goobar14, Zheng14,Tsvetkov14}),
consistent with the expansion in a very clean environment. The same conclusion
is supported by the deep limits on the radio synchrotron emission 
by \cite{Chomiuk14} and \cite{Chandler14}.

\emph{Cessation of mass-loss.} Our observations at $\delta t\sim20$ days 
cannot exclude the presence of material in the environment at larger distances 
$R\gtrsim 10^{16}\,\rm{cm}$ and are thus consistent with any scenario
that predicts  cessation of mass loss from the progenitor system at 
$t\geq300\times(v_{\rm{w}}/10\,\rm{km\,s^{-1}})^{-1}$ yrs before the terminal explosion
("delayed explosion models" in Fig. \ref{Fig:SDphasespace}). This condition
is naturally satisfied by (i) spin-up/spin-down models (\citealt{DiStefano11}; \citealt{Justham11}) 
and (ii) core-degenerate scenarios for Type Ia SNe (where the WD merges with the
core of an AGB -Asymptotic Giant Branch- star, \citealt{Ilkov12}).

In both cases rapid rotation might stabilize the WD against explosion, allowing the WD
to grow above $M_{\rm{Ch}}$. The actual 
delay time $\tau$ between completion of mass transfer and explosion 
is uncertain and depends on the physical mechanism that regulates
the WD spin-down (i.e. reduced accretion, ceased mass transfer, angular momentum transfer, %this from Hachisu 2012
magnetic breaking, gravitational waves). A wide range of values seems to be allowed by theory
($\tau\lesssim 10^5$ yrs to $\tau> 10^9$ yrs, e.g.  \citealt{Lindblom99}, \citealt{Yoon05},
\citealt{Ilkov12}, \citealt{Hachisu12}). Here it is worth noting that  $\tau\geq10^{5}$ yrs
is enough for the circumbinary material to become diffuse and reach ISM-like
density values at the explosion site, consistent with our findings. 
Additionally, for $\tau\gtrsim 10^8-10^9$ yrs, even if the donors started the mass-loss
as giant, sub-giant or main sequence stars, by the time of the explosion 
they have exhausted most of their envelopes, their remaining envelopes have shrunk inside
their Roche lobes and/or they are likely 
to have already evolved into compact objects (e.g. He or C/O WDs) or low-mass stars 
(e.g. \citealt{DiStefano11}, \citealt{Justham11}, \citealt{Hachisu12}). All these effects
would suppress any prominent signature of the mass-donor companion both before (i.e.
in pre-explosion images) and after the explosion (i.e. anything that originates from the
interaction of the SN shock with the medium), naturally accounting for our null-detections.

%In spin-up/spin-down models the WD critical mass needed for explosion
%is increased ($M_{\rm{crit}}>M_{\rm{Ch}}$) as a result of accretion of material 
%with significant angular momentum. If the maximum WD mass is $<M_{\rm{crit}}$,
%an interval $\tau$ of spin down will precede the actual explosion, with   
%$\tau \gg 300$ yrs. Spin-down happens as  a consequence of reduced accretion
%and/or ceased mass transfer. The value of $\tau$ is highly uncertain  and $\tau>10^{9}$ yrs
%is also allowed by theory (..). Here it is worth noting that  $\tau\geq10^{5}$ yrs
%is enough for the circumbinary material to become diffuse and reach ISM-like
%density values at the explosion site, consistent with our findings. 
%Additionally, the donors might have started the mass-loss
%as giant, subgiant or MS stars: however, by the time of the explosion donors
%have exhausted most of their envelopes, their remaining envelopes have shrunk inside
%their Roche lobes and/or  they are likely 
%to have evolved into compact objects (e.g. WDs) or low-mass stars. All these effects
%would diminish our chances of direct detections of the companion star in 
%pre-explosion images.
%No SSS expected to be detected.

%---------------------------------------------------------------------------
\subsection{WD-WD progenitors}

In double-degenerate models the merger of two WDs leads to the final explosion.
The general expectation is that of a ``clean''  environment  at $R\gtrsim 10^{13}-10^{14}\,\rm{cm}$ with ISM-like
density.\footnote{Recent DD studies pointed out the possible presence of a region of enhanced density located at $R\sim10^{13}-10^{14}\,\rm{cm}$
(e.g. \citealt{Fryer10}, \citealt{Shen12}, \citealt{Raskin13}). Its physical origin has to be connected 
either to the WD-WD merger and/or to the outcome of the subsequent
evolution of the system.} 
No stellar progenitor is expected to be detectable at the distance of SN\,2014J
in near UV to NIR pre-explosion images. Additionally, since the WDs were unlikely to burn hydrogen
just prior to explosion,  no X-ray source is expected to be found in pre-explosion images either.
Our deep X-ray non-detection is consistent with this scenario, but can hardly be 
considered a confirmation of this progenitor channel. We also note that our results
are consistent with the predictions of the core-degenerate (CD) scenario, where the
actual merger involves a WD and the core of an asymptotic giant branch star,
as suggested in the case of SN\,2011fe (\citealt{Soker14}).

WD-WD mergers are however expected to enrich the ISM in a number of 
ways including (i) tidal tail ejections (\citealt{Raskin13}), 
(ii) mass outflows during the rapid accretion phase before the merger (\citealt{Guillochon10},
\citealt{Dan11}), 
(iii) winds emanating from the disk during the viscous evolution (\citealt{Ji2013}),  (iv) shell ejections (\citealt{Shen13}. See also \citealt{Soker13}).
Each of these mechanisms predicts the presence of relevant mass at different
distances from the explosion center. Below we discuss the predictions 
of these scenarios at $R\sim10^{16}\,\rm{cm}$ which is relevant for the
\emph{Chandra} null detection of SN\,2014J.

\emph{Tidal tail ejection.} Some $10^{-4}-10^{-2}\,\rm{M_{\sun}}$ of material
is expected to be tidally stripped and ejected with typical velocity $v_{\rm{ej}}\sim 2000\,\rm{km\,s^{-1}}$
just prior to the WD-WD coalescence (\citealt{Raskin13}). 
The resulting structure of the CSM and the
distance $R_{\rm{ej}}$ of the over-density region created by the mass ejection from the explosion center critically
depend on the delay time between the tidal tail ejection and the final explosion $\Delta t_{\rm{ej}}$.
In particular, the ejected material would require $\Delta t_{\rm{ej}}\sim 10^{8}\,\rm{s}$
to reach the distance of interest $R_{\rm{ej}}\sim 10^{16}\,\rm{cm}$.\footnote{Note that for typical
ISM densities $n_{\rm{ISM}}\sim1\,\rm{cm^{-3}}$, the ejected material is in free expansion for $\sim200$ yrs before
entering the Sedov-Taylor phase.} \citealt{Raskin13}
calculate that the tidal tail
density at this distance would correspond to an effective mass-loss rate 
$\dot M\sim 10^{-2}-10^{-5}\,\rm{M_{\sun}\,yr^{-1}}$ which is ruled out by our observations.
%(\textbf{Ref to Fig if you decide to include that in the plot}).
Our results thus argue against WD-WD mergers with delay times of the order of 1-10 yrs,
but allow for short delay times $\Delta t_{\rm{ej}}<10^{6}\,\rm{s}$ (including systems
that detonate on the dynamical time scale $\Delta t_{\rm{ej}}\sim10^2-10^3\,\rm{s}$
or on the viscous time scale of the remnant disk $\Delta t_{\rm{ej}}\sim10^{4}-10^{8}\,\rm{s}$).
Long lag times $\Delta t_{\rm{ej}}\gg10^{8}\,\rm{s}$ are also consistent with our findings,
since the interacting material would be located at larger distances.

\emph{Mass outflows during rapid accretion.} \cite{Guillochon10} and \cite{Dan11} 
suggested that a phase of mass transfer
and rapid accretion (with accretion rates reaching $10^{-5}-10^{-3}\,\rm{M_{\sun}}s^{-1}$) 
might precede the actual WD-WD merger. As for SD systems, mass can be lost at the Lagrangian
point, thus shaping the environment that the SN shock will probe later on.
According to the simulations by \cite{Dan11}, by the time of the merger the
mass of the unbound material is $M_{\rm{ej}}\sim10^{-2}-10^{-3}\,\rm{M_{\sun}}$ and could
in principle produce observable signatures as soon as the SN shock and ejecta expand and interact
with this material. The dynamics of the ejected material depends on $v_{\rm{ej}}$,
$M_{\rm{ej}}$ and $n_{\rm{ISM}}$. The location of the ejected material $R_{\rm{ej}}$ at the time
of the SN explosion critically depends on $\Delta t_{\rm{ej}}$. Both  $\Delta t_{\rm{ej}}$ and
$v_{\rm{ej}}$ are at the moment poorly constrained. For  $v_{\rm{ej}}\sim$a few 
$1000\,\rm{km\,s^{-1}}$ (i.e. comparable to the escape velocity), our deep density limit 
at $R\sim10^{16}\,\rm{cm}$ implies that the system did not eject substantial material 
in the 2-3 years before the SN explosion.

\emph{Disk winds.} WD-WD mergers that fail to promptly detonate 
have been recently suggested to produce a rapidly rotating, magnetized 
WD merger surrounded by a hot thick accretion disk before producing a
Type Ia SN (\citealt{Ji2013}). A fraction of the disk mass is lost through
magnetically driven winds. \cite{Ji2013} calculate that for near-equal 
mass WD, $M_{\rm{ej}}\sim10^{-3}
\,\rm{M_{\sun}}$ is gravitationally unbound and ejected with 
$v_{\rm{ej}}\sim2600\,\rm{km\,s^{-1}}$. As for the other scenarios, a
critical parameter is the interval of time between the mass ejection and the
supernova explosion $\Delta t_{\rm{ej}}$ (assuming that a supernova occurs).
The low density of SN\,2014J at $R\sim10^{16}\,\rm{cm}$ suggests
$\Delta t_{\rm{ej}}>$ a few yrs.

\emph{Shell ejection.} In the case  of He+C/O WD progenitors in the context of the ``double detonation'' model
(\citealt{Livne90}), \cite{Shen13} have recently proposed that hydrogen-rich material can be ejected
by the binary system in multiple ejection episodes hundreds to thousands of years prior to the merger
(see their Fig. 3).
In their simulations \cite{Shen13} find that a total of $M_{\rm{ej}}=(3-6)\times 10^{-5}\,\rm{M_{\sun}}$
of material can be transferred to the environment and it is ejected with initial velocity $v_{\rm{ej}}\sim1500
\,\rm{km\,s^{-1}}$ (i.e. comparable with the He WD circular velocity). In close analogy to nova
shells, the ejected material interacts with the local ISM and shapes the local environment 
of the explosion (\citealt{Shen13}, their Fig. 4). Both the delay time between the onset of the
mass-transfer and the disruption of the He WD and the time of the last ejection 
critically depend on the evolutionary history of the He WD, with the ``older''  He WD in the simulations
of \citealt{Shen13} giving origin to mass ejections $\sim 200$ yrs before the terminal explosion.
Our finding of a very low density environment at $R\sim 10^{16}\,\rm{cm}$
constrains the most recent mass ejection in SN\,2014J (if any) to have happened at 
$t>3$ yrs before the explosion.
%---------------------------------------------------------------------------
%%%%%%%%%%%%%%%%%%%%%%%%%%%%%%%%%%%
%secondary star
%companion star
%donor star
%mass-donor companion
%%%%%%%%%%%%%%%%%%%%%%%%%%%%%%%%%%%%%%%%%%%
\section{Conclusions}
\label{Sec:conclusions}

Type Ia SNe remain among the class of SNe that still lacks both an X-ray and radio
counterpart identification.  Our \emph{Chandra} monitoring campaigns of the
two most nearby SNe Ia in the last 30 years, SN\,2011fe \citep{Margutti12},
and SN\,2014J (this work), allowed us to sample a new region of the $L_x$ vs.
time parameter space ($L_x < 7\times 10^{36}\,\rm{erg\,s^{-1}}$, Fig. \ref{Fig:Xrays}), 
and offered us the opportunity to probe the environment density 
down to the unprecedented level of a few particles per $\rm{cm}^{3}$ at 
$R\lesssim 10^{16}\,\rm{cm}$.

Our major findings can be summarized as follows:
\begin{itemize}
\item The deep X-ray null detection of SN\,2014J at $\sim20$ days 
($L_x < 7\times 10^{36}\,\rm{erg\,s^{-1}}$) implies a low density environment
with $n_{\rm{CSM}}<3\,\rm{cm^{-3}}$
at $R\sim 10^{16}\,\rm{cm}$ from the center of the explosion (ISM medium).
For a wind-like density profile, the luminosity above translates into a pre-explosion
mass-loss rate $\dot M<10^{-9}\,\rm{M_{\sun}yr^{-1}}$ for wind velocity $v_{\rm{w}}=100
\,\rm{km\,s^{-1}}$.
By contrast to limits derived from radio synchrotron emission,
our results are independent of assumptions about the 
poorly constrained post-shock energy density in magnetic fields (e.g.
\citealt{Chomiuk12}, \citealt{Horesh12} for SN\,2011fe).
%For example, for the BL-Ic SN2002ap Bjornsson \& Fransson 2004 demonstrated that the
%assumption of equipartition leads to a sever underestimate of the mass-loss rate. Clear deviations
%from equipartition have also been found for Type-IIb SN2011dh (Soderberg 2012, Horesh 2012).

\item These results rule out with high confidence the majority of the parameter space
populated by SD models with steady mass-loss until the terminal explosion (Fig. 
\ref{Fig:SDphasespace}), and in particular symbiotic systems with a giant companion.

\item \emph{If} the progenitor system of SN\,2014J is SD and harbors a main sequence or
subgiant mass-donor star undergoing  Roche lobe overflow, our results
imply a high efficiency of the accretion process onto the WD $\dot M_{\rm{acc}}/\dot M_{\rm{trans}}\gtrsim99$\%.

\item In the allowed portion of the progenitor parameter space (Fig. 
\ref{Fig:SDphasespace}) we find WD-WD scenarios
(both DD with a pair of C/O WDs and He+C/O WDs of the double detonation model),
SD systems undergoing unstable hydrogen burning, and SD or DD models where the terminal
explosion is delayed with respect to the completion of the mass loss.
A delay time as short as $\sim300\times (v/10\,\rm{km\,s^{-1}})^{-1}$ yrs 
would be enough to explain our X-ray null-detection.
Alternatively, SD progenitors burning hydrogen unsteadily experience repeated nova shell ejections.
\emph{If} SN\,2014J originates from this class of progenitors, our results suggest that,
for typical nova-shell and SN environment parameters, 
(i) at least one nova shell has been ejected at $t\geq40$ yrs before the SN, (ii)
the recurrence time of shell ejection by our system is $<300$ yrs.
 
\end{itemize}

This discussion emphasizes that a single deep X-ray observation of a nearby Type Ia SN
has strong discriminating power in the case of progenitors with steady mass-loss
until the terminal explosion. However, a change in the observation strategy, with multiple
deep observations obtained in the first $\sim 100$ days since the explosion is needed
to better constrain scenarios with sporadic mass-loss episodes just before the terminal
explosion. We advocate for such 
observations to be obtained for the next Type Ia SN discovered in the local Universe ($d<10\,\rm{Mpc}$).
Late-time ($t\gg100$ days) deep X-ray observations will be able to sample the more 
distant environment: for example, recent analysis of the Type Ia SN remnant Kepler suggests that CSM cavities may reach out
to $R\sim10^{17}\,\rm{cm}$ \citep{Patnaude12}.
This effort must be paralleled by realistic theoretical predictions of the environment
density structure around the explosion, possibly shaped by repeated mass ejections from the progenitor
system for a wide range of progenitor/environment parameters.\\
 
%%%%%%%%%%%%%%%%%%%%%%%%%%%%%%%%%%%%%%%%%%%%%%%%%

\acknowledgments 
We thank H. Tananbaum and the entire \emph{Chandra} team for making the 
X-ray observations possible. R. M. thanks Lorenzo Sironi, Cristiano Guidorzi and James
Guillochon for many 
instructive discussions and Georgios Dimitriadis for clarifications about his nova ejection simulations. 
Support for this work was provided by the David 
and Lucile Packard Foundation Fellowship for Science and Engineering awarded to A. M. S.
R. K. acknowledges support from the National Science Foundation through grant AST12-11196.

\bibliographystyle{apj}
%\bibliography{margutti}
%journals_apj

\end{document}